\documentclass{aip-cp}
\usepackage{savesym}
\savesymbol{iint}
\savesymbol{iiint}
\savesymbol{iiint}
\savesymbol{iiiint}
 \savesymbol{idotsint}
\usepackage{amsmath}

\usepackage[numbers]{natbib}
\usepackage{graphicx}
\usepackage{aas_macros}

\begin{document}

\title{Unveiling the Magnetic Structure of VHE SNRs/PWNe with XIPE, the X-ray Imaging-Polarimetry Explorer}

\author[aff1]{E. de Ona Wilhelmi\corref{cor1}}
\author[aff2]{J. Vink}
\author[aff3]{A. Bykov}
\author[aff4]{R. Zanin}
\author[aff5,aff6,aff7]{N. Bucciantini}
\author[aff5]{E. Amato}
\author[aff5]{R. Bandiera}
\author[aff5,aff6,aff7]{B. Olmi}
\author[aff3]{Yu.Uvarov}
\author{for the XIPE Science Working Group}

\affil[aff1]{Institute of Space Sciences (IEEC--CSIC), Carrer de Can Magrans S/N, 08193 Barcelona, Spain.}
\affil[aff2]{Anton Pannekoek Institute for Astronomy \& GRAPPA, University of Amsterdam, Science Park 904, 1098 XH Amsterdam, The Netherlands,}
\affil[aff3]{A. F. Ioffe Institute for Physics and Technology, St. Petersburg, 194021, Russia.}
\affil[aff4]{Max-Planck-Institut fur Kernphysik, P.O. Box 103980, D 69029 Heidelberg, Germany.}
\affil[aff5]{INAF - Osservatorio Astrofisico di Arcetri, Largo E. Fermi 5, I-50125 Firenze, Italy.}
\affil[aff6]{INFN - Sezione di Firenze, Via G. Sansone 1, I-50019 Sesto F. no Firenze, Italy.}
\affil[aff7]{Dipartimento di Fisica e Astronomia, Università degli Studi di Firenze, Via G. Sansone 1, 50019 Sesto F.no (Firenze), Italy}

\corresp[cor1]{Corresponding author: wilhelmi@ice.csic.es}

\maketitle

\begin{abstract}
The dynamics, energetics and evolution of pulsar wind nebulae (PWNe) and supernova remnants (SNRs), are strongly affected by their magnetic field strength and distribution. They are usually strong, extended, sources of non-thermal X-ray radiation, producing intrinsically polarised radiation. The energetic wind around pulsars produces a highly-magnetised, structured flow, often displaying a jet and a torus and different features (i.e. wisps, knots). This magnetic-dominant wind evolves as it moves away from the pulsar magnetosphere to the surrounding large-scale nebula, becoming kinetic-dominant. Basic aspects such how this conversion is produced, or how the jets and torus are formed, as well as the level of turbulence in the nebula are still unknown. Likewise, the processes ruling the acceleration of particles in shell-like SNRs up to 10$^{15}$ eV, including the amplification of the magnetic field, are not clear yet. Imaging polarimetry in this regard is crucial to localise the regions of shock acceleration and to measure the strength and the orientation of the magnetic field at these emission sites. X-ray polarimetry with the X-ray Imaging Polarimetry Explorer (XIPE) will allow the understanding of the magnetic field structure and intensity on different regions in SNRs and PWNe, helping to unveil long-standing questions such as i.e. acceleration of cosmic rays in SNRs or magnetic-to-kinetic energy transfer. SNRs and PWNe also represent the largest population of Galactic very-high energy gamma-ray sources, therefore the study of their magnetic distribution with XIPE will provide fundamental ingredients on the investigation of those sources at very high energies. We will discuss the physics case related to SNRs and PWNe and the expectations of the XIPE observations of some of the most prominent SNRs and PWNe. 
\end{abstract}

\section{INTRODUCTION}

Acceleration phenomena dominate the energy output of many X-ray sources, and their effects are often felt at very large distances from the acceleration site. Polarimetry at radio, infra-red and optical/UV wavelengths has already proven its worth in many instances, but X-ray polarimetry, which can probe higher energy accelerated particles, is needed to obtain a complete view of these phenomena. In fact, imaging X-ray polarimetry is vital to understand the structure and level of order of the magnetic field in PWNe and SNRs, without which our understanding of the acceleration mechanisms and matter interactions in these sources, the acceleration site of all-pervading cosmic rays, is necessarily incomplete. Strong polarisation fraction (P) up to a theoretical limit of P$\sim$0.75 comes from synchrotron emission originating in ordered magnetic fields. It turns out that in most synchrotron sources, such as PWNe and SNRs, but also jets in $\mu$-quasars and active galactic nuclei (AGN), X-rays are emitted by hot electrons located directly at the plasma acceleration sites. The X-ray polarisation is a unique tracer of the magnetic field at these locations, in particular the imaging technique allows the determination of the distribution of the field pointing directly to the acceleration sites. The XIPE mission would spatially resolve a number of PWNe, SNR adding a dependence of the polarisation fraction and the polarisation position angle ($\Psi$) on the position in the sky. 
\section{XIPE}

XIPE\footnote{http://www.isdc.unige.ch/xipe/} is a new mission concept selected by ESA in June 2015 to undergo a 2 years-long assessment phase in the context of the Cosmic Vision M4 competition. The mission is devoted to the observation of celestial sources in X-rays in the 2 keV to 8 keV energy range, with imaging, timing and polarisation capabilities \cite{2013ExA....36..523S}. XIPE has good angular resolution of less than 30 arcseconds is such that the flux from the weakest sources is, even in the less favorable cases, dominating over the background. A large total collecting area can be achieved with three telescopes of 3.5 m focal length that can be harboured within the fairing of the VEGA launcher, without any deployable device. The relatively wide field of view (15 arcmin $\times$ 15 arcmin) and the photon by photon handling of data defines mild requirements for the attitude control. It has a polarisation sensitivity of 1.2\% MDP for 2$\times10^{-10}$ erg/cm$^2$/s in 300 ksec. Aside from spectral, spatial and timing information on the X-ray intensity, XIPE records two additional Stokes parameters, Q and U, as a function of position, photon energy and time. In this way a much wider set of observables encoded in X-ray radiation is explored and helps to break degeneracies in the X-ray modelling of a wide range of astrophysical objects.  XIPE is currently under phase A study by ESA and will be launched in 2025 if selected. 

\section{Simulating XIPE polarisation images}
The XIPE detectors are gas-filled detectors.  X-ray photon will lead to the ejection of a photo-electron, which subsequently will
cause a cascade of further ionisations, releasing secondary electrons. The electron cloud is recorded, and
from this the ejection direction of the primary photo-electron is determined. The primary electron's direction angle $\phi$ has a $\cos^2(\phi-\psi_0)$ modulation, with $\psi_0$ the electric-field vector (polarisation angle) of the electromagnetic wave. On top of this modulation the measured angle has a also a measurement error, reducing the
modulation amplitude.

The input for our XIPE simulations consists of three images for 1) the morphology of the source (typically a Chandra image), 2) the polarisation
fraction per pixel, and 3) the polarisation angle. The output consists of Monte Carlo realisations of detected photons, with a image position $(i,j)$ and
photo-electron direction ($\phi$). The Monte Carlo realisations take into account the PSF of the telescope \citep[based on][]{fabiani14},
the input polarisation fraction and angle, 
the $\cos^2$ modulation of $\phi$ plus a Gaussian error with $\sigma(\phi)=35^\circ\sqrt{E/4\mathrm{keV}}$, which gives a reasonable match with the measured polarised fraction error of the detector \citep[][Fig.~11]{muleri10}. For the energy distribution of the Monte Carlo events and the count rate, we used the latest estimates for the total effective area of XIPE mirrors plus detectors.

From the resulting Monte Carlo event list Stokes I, Q, and U images  can be made following the procedure explained in \citep{kislat15}. 
For $N_{i,j}$ events for a given image pixel, the estimated values for the  stokes parameters are
\begin{align}\label{eq:qu_est}
I_{i,j} = & \sum_{k=1}^{N_{i,j}} 1 =N_{i,j}, \\\nonumber
Q_{i,j}= &2\sum_{k=1}^{N_{i,j}} \cos^2(\phi_k) - \sin^2(\phi_k)= 2 \sum_{k=1}^{N_{i,j}} \cos (2\phi_k), \\\nonumber
U_{i,j}= & 2 \sum_{k=1}^{N_{i,j}} 2\sin \phi_k \cos \phi_k = 2 \sum_{k=1}^{N_{i,j}}\sin(2\phi_k).
\end{align}
The variances in these estimates ($<Q^2>_{i,j},<U^2>_{i,j}$) can be obtained by summing the squares of the kernels (i.e. respectively $4 \cos^2(2\phi_k),4\sin^2(2\phi_k)$).
These equations look very similar to the definition of Stokes parameters based on the electromagnetic vector angle $\psi$, but an additional factor $2$ is included, to compensate for the fact that not the electric vector angle $\psi$ is measured, but
a realisation of a $\cos^2$ modulation of it \citep{kislat15}. In principle, another, energy dependent, factor can be included $f(E_i)$ in the kernel of $Q$ and $U$
and their variances, in order to compensate for the energy dependent measurement error of $\phi_i$. This was not done in the simulations presented here, but may be in practice useful to factor in the energy dependent error of the photo-electron angle.

\begin{figure}[h]
  \centerline{\includegraphics[width=305pt]{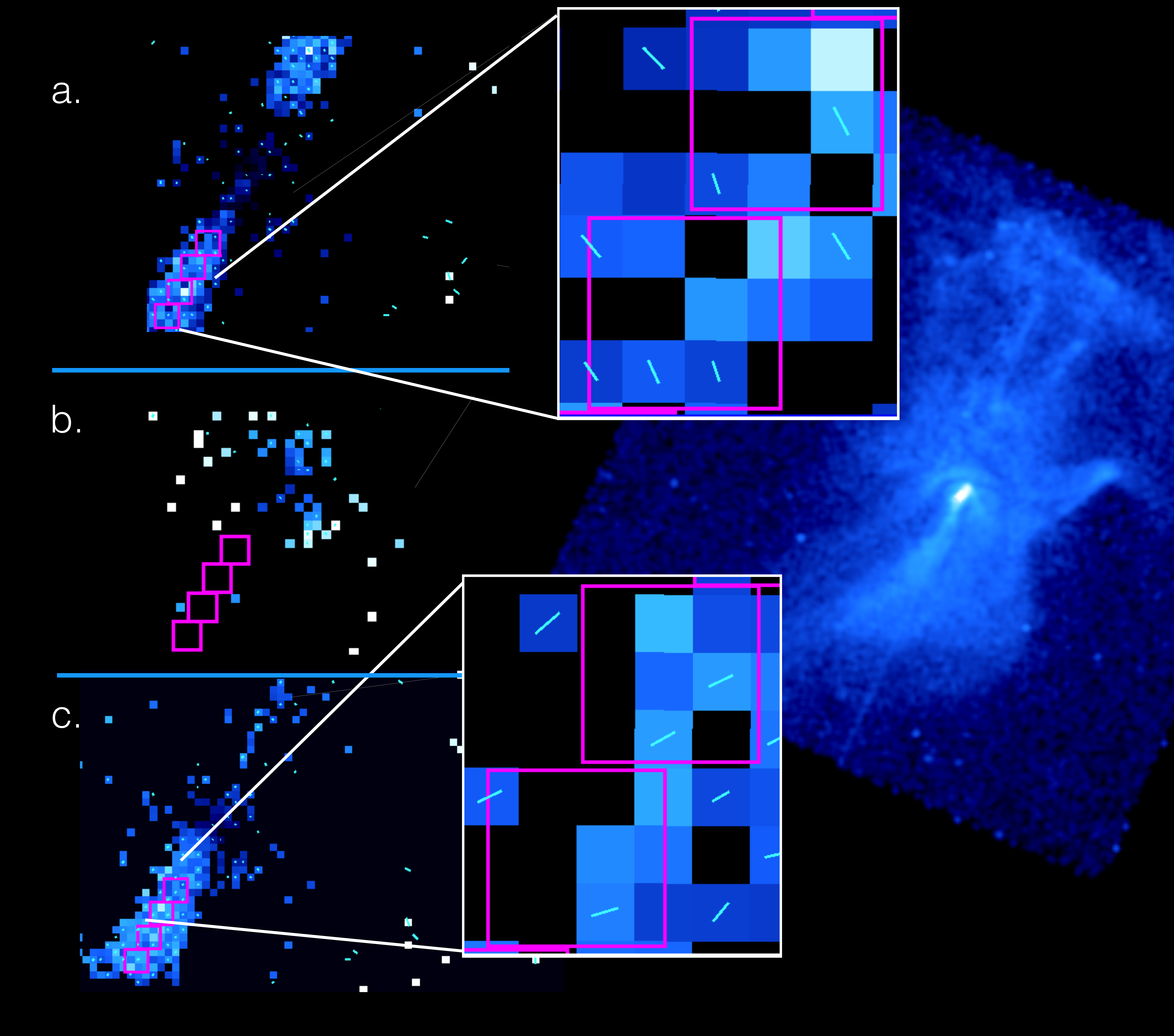}}
  \caption{MSH 15-52 simulations as seen by XIPE in 2 Msec. On the left, the three images (a, b, and c) show the intensity map expected when applying to the Chandra image (on the background) and using the spectrum measured by \cite{2014ApJ...793...90A} for three different models of polarisation. The toy models consist on a jet and torus magnetic field structure with an ordered toroidal component plus a) a fully ordered radial magnetic field, b) a fully disordered magnetic field and c) a fully perpendicular magnetic field. The first and third case are magnified to show the reconstructed direction. }
  \label{fig:1}
\end{figure}

\begin{figure}
  \centerline{\includegraphics[width=305pt]{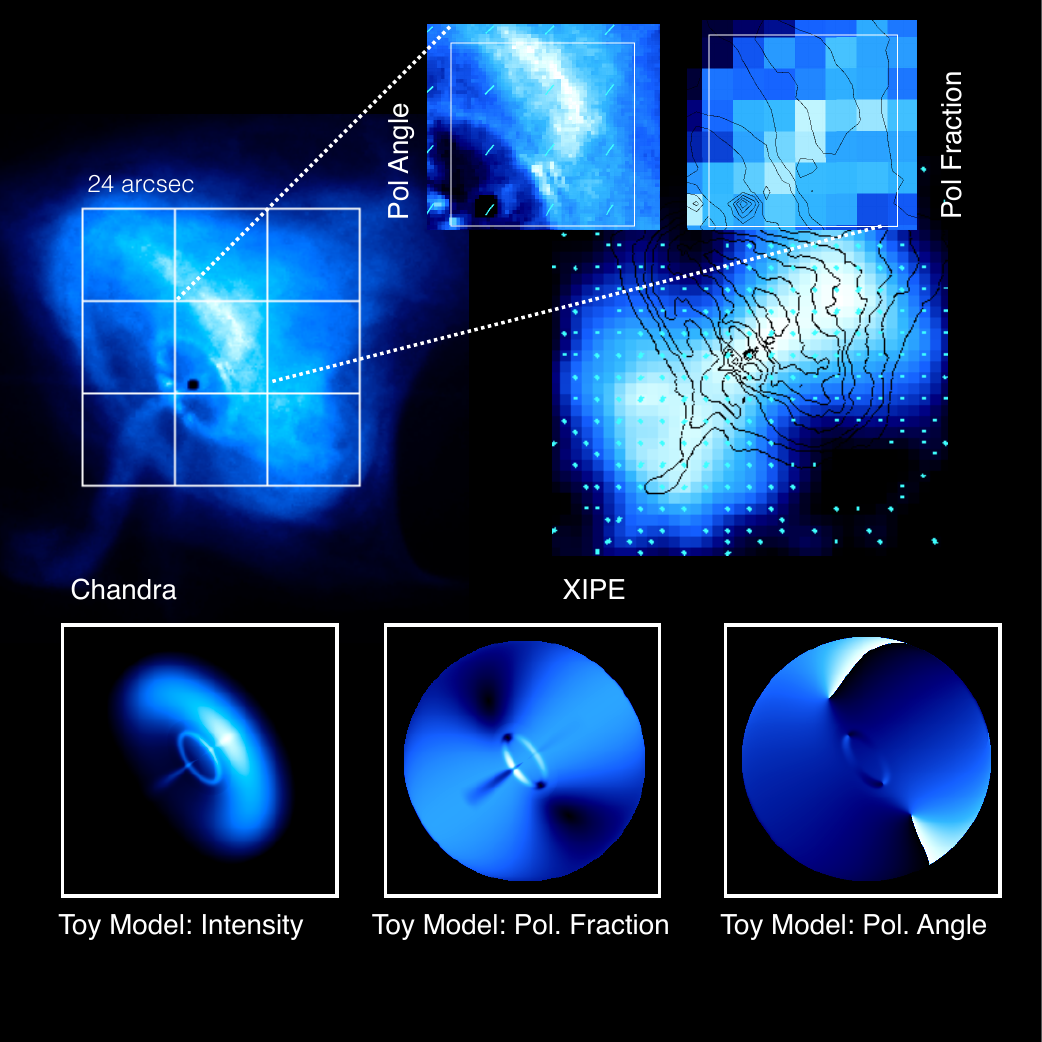}}
  \caption{Crab Nebula simulations as seen by XIPE in 0.2 ksec. The toy model mimic the Chandra image for a given polarisation angle $\Psi$ and fraction P. The reconstructed XIPE intensity image on the right is magnified on the top, to show the optimal direction and intensity reconstruction on a 24$x$24 arcsecond$^2$ region.}
\label{fig:2}
\end{figure}

The polarisation fraction for a given pixel position is  given by
\begin{equation}
P_{i,j} = \frac{\sqrt{{Q_{i,j}}^2+{U_{i,j}}^2}}{I_{i,j}},
\end{equation}
and its variance can be estimated through error propagation from $<Q^2>_{i,j},<U^2>_{i,j}$.
It is often more useful to display $P_{i,j} \times I_\mathrm{i,j}$, the so-called polarised intensity image.
The polarisation angle is given by
\begin{equation}
\psi_{i,j,0}=\frac{1}{2}\arctan\left( \frac{U_{i,j}}{Q_{i,j}}\right).
\end{equation}
The results of these procedures are shown in Fig.~\ref{fig:1}-\ref{fig:3}.


\section{Acceleration process in PWNe and SNRs observed by XIPE}
\subsection{XIPE Prospects for Pulsar Wind Nebulae}

PWNe are bubbles of relativistic particles and magnetic field formed by the interaction of the relativistic Pulsar Wind with the surrounding SNR. They are among the most efficient particle accelerators in our Galaxy \cite{2007MNRAS.381.1489N,2008A&A...485..337V} and emit synchrotron emission in the X-ray band. At present, the Crab Nebula is the only astronomical source with a high-confidence X-ray polarimetric measurement (P=19.2+/-1.0, \cite{1978ApJ...220L.117W}). Recent INTEGRAL results \cite{2008Sci...321.1183D,2008ApJ...688L..29F} also suggest a high level of gamma-ray polarisation.
The magnetic field in PWNe can be probed by optical and radio polarimetry. It is well-ordered with P being as high as 0.6. This is close to the theoretical limit. As laid out in the introduction, the X-rays are produced close to where the electrons are accelerated and therefore provide a much cleaner view of the inner regions than optical observations. The latter also suffer from depolarising foreground effects. Detailed morphological studies with Chandra revealed a complex structure of PWNe, such as, e.g., the torus-plus-jet structure in the Crab Nebula \cite{2000ApJ...536L..81W}. In general, young pulsars show an axial symmetry around what is believed to be the pulsar rotation axis \cite{2002MNRAS.329L..34L,2003MNRAS.344L..93K,2004A&A...421.1063D,2005MNRAS.358..705B}.
Spatially-resolved measurements by XIPE will determine the magnetic field orientation and the level of turbulence in the torus, the jet, and at various distances from the pulsar. This is of special interest because the polarised emission is more sensitive to the plasma dynamics in PWNe than the total synchrotron emission \cite{2005A&A...443..519B}. Knowing how the level of turbulence changes with distance from the shock could test recent MHD scenarios that invoke the conversion of magnetic energy into particle energy inside the radiation region \cite{2014MNRAS.438..278P}. Moreover, comparing X-ray to optical polarisation for the inner bright features could clarify if the sites and mechanism(s) of particle acceleration at different energies are the same.

Indeed, differently from young pulsars, more evolved systems rather show a crushed morphology. It is caused by the interaction between the PWN and the host SNR. Still, a wide diversity in morphology and time variability is found within both classes of young and evolved systems; PWNe around pulsars with similar characteristics may show striking differences with each other. A likely explanation for this relies on the relative orientation between the spin and magnetic dipole axes. If this relative orientation evolves systematically with PWN morphology, this would also reflect a systematic evolution of the polarisation angle and fraction. To demonstrate the capability of XIPE to disentangle different structures and magnetic field distributions, we performed Monte Carlo simulations using the baseline combined telescope effective area and point spread function (PSF). XIPE simulations of three different scenarios (see Figure \ref{fig:1} for a realistic 2Msec simulation of the PWN MSH 15-52) show its outstanding response to different values of (P,$\Psi$). The polarised features applied on the Chandra image \cite{2014ApJ...793...90A}, using the three toy models (by N. Bucciantini) are reconstructed with errors of less than 0.1\% and more than 10$\sigma$ within the instrument PSF. For details on the simulation code  and the generation of $Q$ and $U$ images from XIPE events, see the last section.

In young pulsars, XIPE will unveil the interplay between the pulsar and its relativistic wind. When \cite{1984ApJ...283..710K} published the first MHD description of the plasma flow in the Crab Nebula, the low level of magnetisation of the flow immediately upstream of the pulsar wind termination shock remained unexplained. This is the so-called sigma problem. Recent 3D MHD simulations \cite{2014MNRAS.438..278P} suggest that at the shock front an energy equality between particles and magnetic field is required to reproduce the Crab Nebula morphology; thanks to kink instabilities, the magnetic field would become progressively more tangled and thus dissipates in the bulk of the nebula, i.e. in the emitting region rather than in the non-radiative cold pulsar wind. X-ray polarimetry offers the best possible tool to investigate this scenario, which, if proven to be true, would at the same time solve the sigma problem and provide new insight into dynamics of the relativistic plasma in PWNe. The potencial of XIPE to describe the magnetic structure of the Crab nebula is illustrated in Figure \ref{fig:2}, in which a toy model (by N. Bucciantini) describing the nebula morphology is reconstructed with good accuracy by simulating a 0.2 ksec XIPE observations.  

\begin{figure}[h]
  \includegraphics[width=305pt]{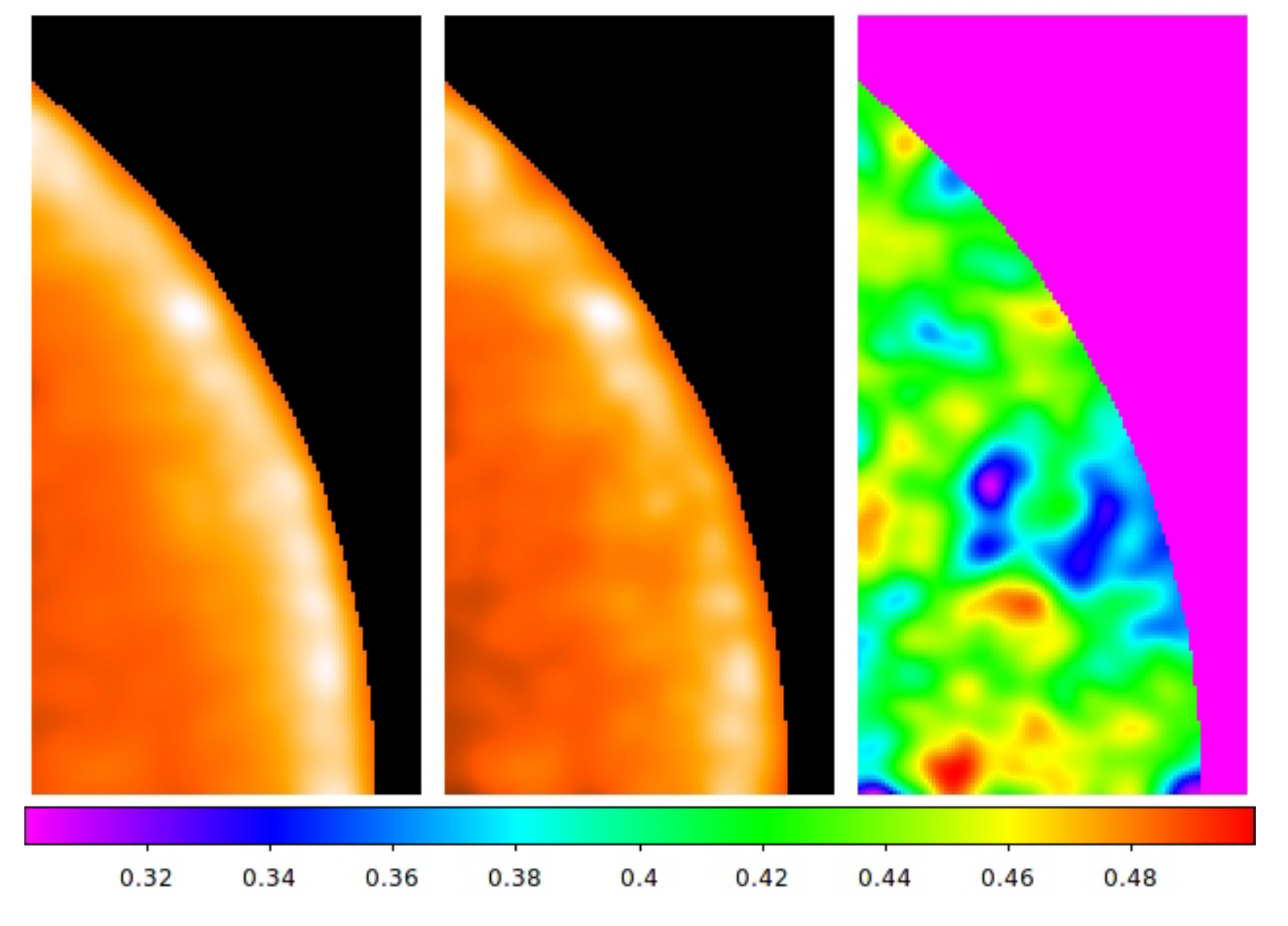}
  \caption{Simulated 4-6 keV maps of polarised synchrotron radiation
  of a supernova shell with turbulent magnetic field. 4-6 keV intensity is shown with a linear
colour scale in the left-hand panel. The right-hand panel shows the
degree of polarisation indicated by the colour bar. The stochastic
magnetic field sample has Kolmogorov type spectral index = 5/3. The
central panel shows the product of intensity and polarisation degree
(see \cite{2009MNRAS.399.1119B}).}
  \label{fig:4}
\end{figure}
\begin{figure}[h]
  \centerline{\includegraphics[width=\textwidth]{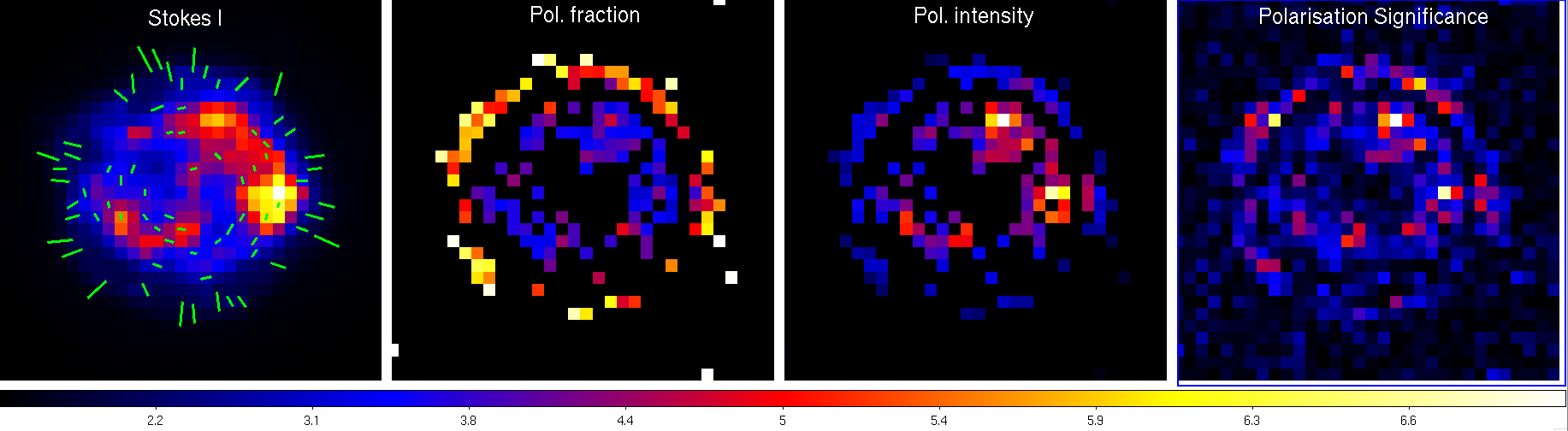}}
  \caption{Monte Carlo simulation of a 4.2-6 keV  4~Ms observation of SNR Cas A with XIPE.
    The simulation is based on the 4.3-6~keV X-ray continuum image  \citep[e.g.][]{hwang04,2008ApJ...686.1094H}.
  The colour bar refers to the polarisation detection significance. Only pixels with significant polarisation detections
  are displayed in the polarised fraction and intensity images.
  It was assumed that only the inner filaments and outer filaments are polarised, hence the ring-like appearance.
  The measured polarisation fraction ranges from $\sim3\%$ for the reverse-shock region to $\sim 7$\% from
  the forward shock region. As expected \citep{muleri10} this is about 50\% of the input values.
  }
\label{fig:3}
\end{figure}

\subsection{XIPE Prospect for Supernova Remnants}

SNRs are relativistic particle accelerators and the most likely sources of Galactic cosmic rays. X-ray synchrotron emitting regions in SNRs have been identified in several sources (see review by \cite{2012SSRv..173..369H}), proving that electrons are accelerated up to 10-100 TeV. At such energies, the time scale of radiation loss is very short so that X-ray synchrotron emission can only be found in regions of active acceleration: the SNR shock fronts.
The X-ray synchrotron emitting regions can be very narrow, their widths being a measure of the average strength of the magnetic field. In some cases, like for Cas A, Tycho and Kepler SNR, the filaments are arc seconds wide and can only be resolved by the Chandra satellite. The filament widths indicate magnetic field strengths of 100--500 mG, well above the Galactic magnetic field strength of $\sim$5 $\mu$G, indicating strong magnetic field amplification \cite{2004MNRAS.353..550B}. Irrespective of the magnetic field strengths, X-ray synchrotron emission can only occur for fast shocks ($>$3000 km/s), and requires small mean free paths for the relativistic electrons, and therefore very turbulent magnetic fields.

XIPE can explore the turbulence level of the magnetic field, which plays a crucial role in diffusive shock
acceleration theories, and determines whether protons, accelerated under the same conditions as the
electrons, can be accelerated to energies beyond 100 TeV \cite{2009MNRAS.399.1119B}. The latter is needed to explain the Galactic cosmic ray spectrum observed on Earth (e.g. \cite{2014NuPhS.256....9B}) as well as the physical nature of complex plasma processes in collisionless shocks which otherwise can not be studied in laboratories \cite{2016RPPh...79d6901M}. In
Figure~\ref{fig:4} a simulated 4 keV to 6 keV X-ray image of a supernova shell is presented. The model used is described in detail in \cite{2009MNRAS.399.1119B,2011ApJ...735L..40B}. It accounts for anisotropic cosmic ray driven magnetic turbulence in the synchrotron emitting supernova shell with parameters which resemble Tycho's SNR. It is important to note that the X-ray synchrotron emission is coming
from a region within a diffusion length scale from the shock acceleration site, unlike the radio synchrotron emission which originates from the entire shell of the remnant. As a result, the X-ray polarisation is much less affected by depolarisation due to different polarisation angles along the line of sight than in the radio, but on the other hand, the magnetic field is expected to be  more turbulent near the shock front. 
It is, therefore, difficult to asses a priori whether the polarisation fraction in X-rays will be lower or higher than in the radio. Young supernova remnants have been measured  to have low average polarisations in the radio in general, ranging from $\sim 5$\% in Cas A \citep{anderson95b} to 17\% in SN\,1006 \citep{reynoso13}.

To test the ability of XIPE to image the narrow filaments seen in the 4.1-6.1 keV continuum band of bright SNRs such as Cas A, we  performed Monte Carlo simulations (Fig.~\ref{fig:3}, see next section for details). In Cas A there are also interior filaments associated with the reverse shock \cite{2008ApJ...686.1094H}. As an educated guess, we assumed a polarisation fraction of 20\% for the outer shock filaments and of 4\% for the interior continuum emission. To assess the effects of blending, due to the PSF, we assigned a tangential polarisation angle to the interior and a radial orientation to the outer filaments. Remarkably, although the narrow outer rim cannot be resolved with XIPE, the images with polarisation fractions/polarised intensity show that the rims can be detected thanks to its distinct polarisation pattern. The interior has a lower polarisation degree in the simulations, but due to the higher brightness polarisation can be detected also here (see the image with polarisation detection significance).

\section{Conclusions}

Simulations probe that XIPE would represent a step forward in our knowledge of the physics of PWNe and SNRs. In summary, XIPE would:
\begin{itemize}
\item constrain the plasma dynamics inside different subcomponents of several young PWNe,
\item verify if the relative orientation between rotational and magnetic axes of pulsars correlates with the morphology of PWNe,
\item test recent MHD models potentially solving the long-standing sigma problem.
\item image in spatial detail the turbulence level of magnetic fields in SNRs,
\item thereby constraining theories of diffusive shock acceleration in SNRs and their contribution to the Galactic cosmic-ray spectrum and energy density.

\end{itemize}


\section{ACKNOWLEDGMENTS}
E. dO.W. acknowledges the grants AYA2015-71042-P, and SGR2012--1073.  A.B. and Yu.U. acknowledge support from RSF grant 16-12-10225. B. O. acknowledge support from the INFN - TEONGRAV initiative (local PI: Luca Del Zanna) and from the University of Florence grant \emph{Fisica dei plasmi relativistici: teoria e applicazioni moderne}. R.Z. acknowledges the Alexander von Humboldt Foundation for the financial support and the Max-Planck Institut fur Kernphysik as hosting institution.

\nocite{*}
%

\end{document}